\documentclass[aps,groupedaddress,superscriptaddress,amsmath,amssymb,pra,twocolumn,longbibliography]{revtex4-2}
\usepackage{bm}
\usepackage{hyperref}
\usepackage{float}
\usepackage{graphicx}
\usepackage{verbatim}
\usepackage{graphicx}
\usepackage{amsfonts}
\usepackage{listings}
\usepackage[pdftex]{color}
\usepackage{verbatim}
\usepackage{bbm}
\usepackage{braket}
\usepackage{qcircuit}
\usepackage[normalem]{ulem}
\hypersetup{
    colorlinks=true,
    linkcolor=blue,
    filecolor=magenta,      
    urlcolor=cyan,
}
\usepackage{color}

\def\omq{\delta\omega_\mathrm{q}}
\def\sn{^{(n)}}

\def\th{\theta{}}
\def\ts{t_\mathrm{cyc}}
\def\ep{\varepsilon}

\def\mE{\mathbb{E}}
\def\mS{\mathbb{S}}

\def\h1{\mathds{1}}

\begin{document}
\title{Revealing inadvertent periodic modulation of qubit frequency}
\author{Filip Wudarski}
\affiliation{USRA Research Institute for Advanced Computer Science (RIACS), Mountain View, CA 94043, USA}
\author{Yaxing Zhang}
\affiliation{Google Quantum AI, Santa Barbara, CA 93111, USA}
\author{Juan Atalaya}
\affiliation{Google Quantum AI, Santa Barbara, CA 93111, USA}
\author{M. I. Dykman}
\affiliation{Department of Physics and Astronomy, Michigan State University, East Lansing, MI 48824, USA}

\begin{abstract} 
The paper describes the means to reveal and characterize slow periodic modulation of qubit frequency. Such modulation can come from different sources and can impact qubit stability. We show that the modulation leads to very sharp peaks in the power spectrum of  outcomes of periodically repeated Ramsey measurements. The positions and shapes of the peaks allow finding both the frequency and the amplitude of the modulation. We also explore how additional slow fluctuations of the qubit frequency and fluctuations of the modulation frequency affect the spectrum. The analytical results are in excellent agreement with extensive simulations.
\end{abstract}

\date{\today}

\maketitle

\section{Introduction}
\label{sec:Intro}
Progress in quantum computing requires further improving physical qubits, the building blocks of quantum computers. A major obstacle is qubit decoherence, and in particular fluctuations of qubit frequencies. Overcoming it entails understanding the fluctuation mechanisms. These mechanisms have been attracting significant attention. The primary means to reveal them have been the analysis of the fluctuation spectra \cite{Nakamura2002,Bialczak2007,Alvarez2011,Bylander2011,Sank2012,Yan2012,Young2012,Paz-Silva2014,Yoshihara2014,Kim2015,Brownnutt2015,OMalley2015,Yan2016,Myers2017,Quintana2017,Paz-Silva2017,Ferrie2018,Noel2019,vonLupke2020,Proctor2020,Wolfowicz2021,Wang2021,Burgardt2023,Khan2024} and, to a lesser extent, of the fluctuation statistics \cite{Li2013a,Ramon2015,Norris2016,Sinitsyn2016,Szankowski2017,Sung2019,Ramon2019,Sakuldee2020,Wang2020b,You2021,Wudarski2023,Wudarski2023a,Wang2024}. 

Of special interest are low-frequency qubit fluctuations, with frequencies smaller than the decay rates and the  characteristic decoherence rates of the qubits. Such fluctuations control the reproducibility of the results obtained by repeatedly running a quantum code, since they lead to the parameters of the qubits being different from run to run. A major source of low-frequency fluctuations in qubits is believed to be two-level systems (TLSs) \cite{Paladino2002,Galperin2004,Faoro2004,Galperin2006a,Paladino2014,Ramon2015,Muller2019,Ramon2019,Huang2022,Mehmandoost2024}. This is because the fluctuations come from switching between the states of TLSs, and since the switching rates can be very low, the fluctuations spectrum extends to low frequencies. 

Along with TLSs there is another important source of low-frequency perturbations of qubits, which is often disregarded. Many components involved in the functioning of quantum computing hardware, such as cooling systems and vacuum pumps, for example, work in cycles. The voltage from the AC power supplies is  also periodic. The associated periodic perturbations may cause inadvertent periodic modulation of the qubit frequencies. The typical modulation frequencies are in the tens to a few hundred Hertz, or maybe few kilohertz, they are well below the relaxation rates and well below the  spectral range of qubit fluctuations most frequently studied in the experiments. Yet, if the qubit parameters vary at such frequencies, it would affect the performance of a quantum computer. 

In this work we propose a way to reveal and characterize periodic modulation of qubit frequency. The method  circumvents the constraint that comes from the frequency being much smaller than the reciprocal qubit coherence time. It relies on periodically repeating qubit measurements, with the duration of an individual measurement being shorter than the qubit coherence time, but the overall measurement duration being much longer than the modulation period, cf. Fig.~\ref{fig:pulse_sequence}.  A similar sequence of measurements  was used in \cite{Wudarski2023,Wudarski2023a} to reveal the statistics and some characteristic features of random low-frequency qubit fluctuations, including  those induced by TLSs.

An individual short quantum measurement sketched in Fig.~\ref{fig:pulse_sequence} is a measurement of  the phase acquired by a qubit over the measurement duration. This phase is determined by the qubit frequency. If the period of the frequency modulation is much longer than the measurement duration, the frequency remains nearly constant and the phase is proportional to the instantaneous frequency value. If the measurements are repeated with period much shorter than the modulation period, the phase varies slightly between consecutive  measurements. However,  it is different for measurements separated by a significant portion of the modulation period, and it varies periodically with this period. It is this periodicity that our method reveals. As we show analytically and by numerical simulations, the  outcomes of the measurement sequence allow one not only to uncover periodic modulation of the qubit frequency, but also to find the frequency and amplitude of the modulation. 

Besides the periodic change of the qubit frequency the analysis of low-frequency qubit fluctuations should address several additional issues. First, the amplitude of the periodic modulation is not necessarily small, making  it  necessary to understand what determines its actual strength and how the associated nonlinearity is manifested in the measurements.  Moreover, periodic  modulation of the qubit frequency generally comes along with the modulation from random sources. Since qubits are strongly nonlinear systems, the effects of periodic and random frequency modulation on the measurement outcomes are not described by a sum of the effects of each of them and actually result in somewhat unexpected interference-type manifestations. These manifestations are different depending on the noise source. In addition, the ``periodic'' frequency modulation is not strictly periodic in real systems. For example, even in the power grid, the frequency fluctuates \cite{Wood2013,Schafer2018} (the targeted variation is within $\pm 0.5\%$ in the US). The fluctuations may well be larger for mechanical devices. The effect of such fluctuations is an issue that we also address.

The paper is organized as follows. In Section~\ref{sec:correlation_general} we describe the quantum system under investigation: a qubit with a frequency that is periodically modulated and is also modulated by the noise from the environment. We also describe how the measurements are envisioned. Section \ref{sec:general} provides general expressions for the power spectrum of the measurement outcomes in the cases where the measurements are synchronized with the modulation and in a more important case of non-synchronized measurements. In Section \ref{sec:resonant_peaks} we describe a theory and provide the results of simulations of  resonant peaks in the power spectrum and show that they can be used to characterize the modulation frequency and amplitude. In Sec.~\ref{sec:general_frequency_noise} we consider the combined effect of the random and periodic  modulation of the qubit frequency, including the random modulation that comes from the coupling to two-level systems.  Section~\ref{sec:peak_broadening} describes broadening of the spectral peaks due to fluctuations of the modulation frequency or the measurement period. Appendices describe the simulation algorithms and also a method, which differs from the conventional Fourier transform and provides an alternative way of finding the modulation frequency.


\section{Measurement setup}
\label{sec:correlation_general}

Our analysis is based on studying the outcome of the periodically repeated Ramsey measurements of a qubit \cite{Sank2012,Yan2012}. We are interested in this outcome in the case where the qubit frequency $\omega_q$ is periodically or quasi-periodically  modulated and can also fluctuate because of the noise from the qubit environment. To begin, we describe the measurement routine. To this end, we associate the operators acting on the qubit states with the Pauli operators $\sigma_{x,y,z}$ and the unit operator $\hat I_q$.  The ground $\ket{0}$ and excited $\ket{1}$ states of the qubit are the eigenstates of $\sigma_z$. In a Ramsey measurement a qubit, initially in the state $\ket{0}$, is rotated by the transformation $\exp(-i\pi\sigma_y/4)$ into the state $(\ket{0}+\ket{1})/\sqrt{2}$. After time $t_R$, the same unitary transformation is applied again and the occupation of the state $\ket{1}$ is measured.  After the measurement the qubit is reset to the ground state. In our setup, the Ramsey measurements  are  repeated with period $\ts$, which we call the cycle period. 

\begin{figure}[h]
\includegraphics[width=0.47\textwidth]{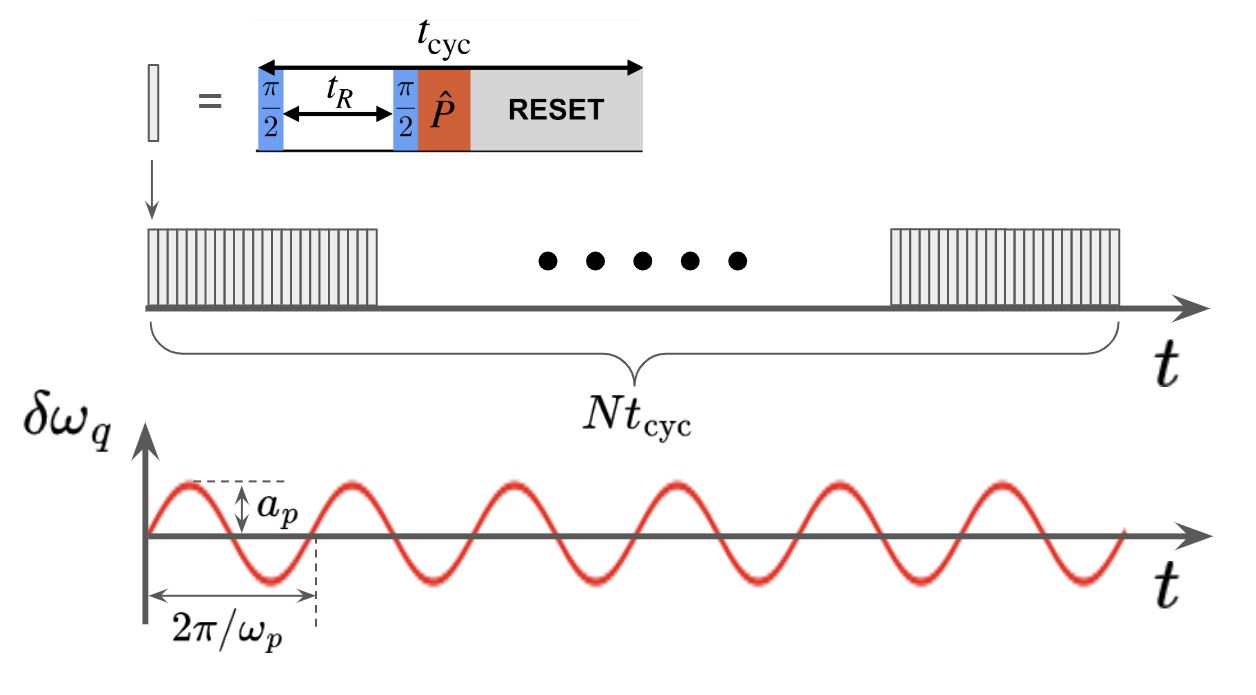} \\
\caption{Schematics of the measurements. An individual measurement is of the Ramsey type. It involves two rotations by $\pi/2$ of the qubit state  on the Bloch sphere, which are separated by time $t_R$ and are followed by a projective measurement $\hat P = \ket{1}\bra{1}$ and a reset to the ground state $\ket{0}$. The measurement outcome is determined by the phase accumulated over $t_R$. The measurements are repeated with period $\ts$. Sketched at the bottom is the periodic addition $\delta\omega_q$ to  the qubit frequency as a function of time}.
\label{fig:pulse_sequence}
\end{figure}

The qubit phase $\theta$ accumulated over time $t_R$ is counted off from the phase accumulated over that time by a signal at the reference frequency. It is thus determined by the deviation $\omq(t)$ of the qubit frequency from this reference frequency. 

For a given $\theta$, the probability of obtaining ``1'' as an outcome of the  Ramsey measurement is \cite{Nielsen2011}
\begin{align}
\label{eq:standard_probability}
p(\theta) = \frac{1}{2}\left[1+e^{-t_R/T_2}\cos(\phi_R + \theta)\right]\ .
\end{align}
Here  $T_2^{-1}$ is the qubit decoherence  rate due to fast decay and dephasing processes. The phase $\phi_R$ mimics the phase accumulated due to the difference between the nominal, i.e., unperturbed, qubit frequency and the frequency of the reference signal. This phase can be (and frequently is)  added in a controlled way by a gate operation.

In what follows we will assume that $t_R$ is small compared to $T_2$ and will disregard the factor $\exp(-t_R/T_2)$ in Eq.~(\ref{eq:standard_probability}). Incorporating this factor is straightforward and does not change the qualitative results.

In the repeated measurements sketched in Fig.~\ref{fig:pulse_sequence}, the  phase $\th_k$ accumulated in the $k$th  cycle, i.e., in  the time interval $(k\ts, k\ts+t_R)$, is 
\begin{align}
\label{eq:phase_k_classical}
\theta_k = \int_{k\ts}^{k\ts+t_R}\omq (t)dt\ .
\end{align}
It varies from cycle to cycle. This variation determines the outcome of the measurements and thus provides a means for revealing the variations of the qubit frequency. .

For a sinusoidal modulation at frequency $\omega_p$, the modulation-induced change of the qubit frequency is  
\[\omq^{(p)}(t) = a_p\cos(\omega_pt+\phi_p).\] 
Here $a_p$ is the modulation amplitude. The phase $\phi_p$ is counted off from the phase of the modulation at time $t=0$ when the measurements start. 

The modulation-induced phase accumulation during the $k$th Ramsey measurement is
\begin{align}
\label{eq:frequency_sinusoidal}
&\theta_k^{(p)}= \int_{k\ts}^{k\ts +t_R}\omq^{(p)}(t) dt=A_p\cos(k \omega_p\ts +\tilde \phi_p)\ ,
\nonumber\\
&A_p= \frac{2a_p}{\omega_p}\sin\frac{\omega_p t_R}{2},\quad \tilde\phi_p=\phi_p+\frac{\omega_p t_R}{2}.
\end{align}
The phase $\theta_k^{(p)}$ is periodic in the cycle number $k$, which is a consequence of the periodicity of the frequency modulation. In the important case of slow modulation compared to the qubit decoherence rate we have $\omega_p T_2\ll 1$, and thus $\omega_p t_R\ll 1$. Then the effective amplitude of the phase oscillations $A_p \approx a_pt_R$ is independent of $\omega_p$, but is proportional to the duration of the Ramsey measurement $t_R$.

The total phase accumulated during the $k$th measurement in the presence of fluctuations is  
\begin{align}
\label{eq:phase_total}
\theta_k = \theta_k^{(p)} +\theta_k^{(r)},\quad \theta_k^{(r)} = \int_{k\ts}^{k\ts + t_R} dt \,\omq^{(r)}(t).
\end{align}
Here $\theta_k^{(r)}$ is the random part of the phase that comes from the noise that drives the qubit and adds a random term $\omq^{(r)}$ to its frequency. We will assume  $\langle\theta_k^{(r)}\rangle = 0$.

 \section{Power spectrum of the measurement outcomes}
 \label{sec:general}

Periodic  modulation of the qubit frequency can be revealed by studying the power spectrum of the Ramsey measurement outcomes $x_n$. These outcomes take on values ``0'' or ``1''.  Their power spectrum for  $N$ measurements is determined by the discrete Fourier transform $X(m)$, 
\begin{align}
\label{eq:fast_Fourier_general}
X(m) = N^{-1/2}\sum_{n=0}^{N-1}x_n\exp(2\pi i mn/N).
\end{align}

The power spectrum is given by the expectation value $\mS(m)=\mE[|X(m)|^2]$. To calculate it we have to take into account that, for the considered quantum measurements, $x_m$ is either 0 or 1, so that $x_m^2 = x_m$. Therefore we have 
\begin{align}
    \label{eq:x_n_x_m}
\mE[x_nx_m] = \braket{p(\theta_n) p(\theta_m)} + \left(\braket{p(\theta_n)} - \braket{p(\theta_n)}^2\right)\delta_{mn}.
\end{align}
where $p(\theta)$ is given by Eq.~(\ref{eq:standard_probability}) while $\langle\cdot\rangle$ indicates averaging over realizations of the noise that drives the qubit.  

At this point we are considering measurements that are ``synchronized'' with the periodic modulation, i.e., the modulation phase $\phi_p$ is the same $(\mathrm{mod}\, 2\pi)$ in each series of $N$ measurements.  With the account of the correlations imposed by the classical noise in $\theta_k$, for such synchronized measurements we have
\[\mS(m) = \mS_\mathrm{syn}(m) + \mS_\mathrm{wht},
\]
where
\begin{align}
\label{eq:power_synchron}
\mS_\mathrm{syn}(m) &= \frac{1}{4N}\sum_{n_1,n_2=0}^{N-1}
\braket{\cos(\theta_{n_1}+\phi_R)\cos(\theta_{n_2}+\phi_R)}
\nonumber\\
&\times \exp[2\pi i m(n_1 - n_2)/N] 
\end{align}
and
\begin{align}
    \label{eq:white_noise}
    \mS_\mathrm{wht} = \frac{1}{8N}\sum_n\left(1-\braket{\cos(2\theta_n+ 2 \phi_R}\right)
\end{align}
The terms $\mS_\mathrm{syn}(m)$ and $\mS_\mathrm{wht}$ come from the first and the second terms in the right-hand side of Eq.~(\ref{eq:x_n_x_m}), respectively. We have used that only the product of the cosines in $\braket{p(\theta_{n_1})p(\theta_{n_2})}$ contributes to the power spectrum for $m\neq 0$. Further, in calculating this product we will use that, for stationary noise,
\begin{align}
\label{eq:phase_average}
\langle \exp[i\theta_n^{(r)} \pm i\theta_m^{(r)}]\rangle = F_\pm (n-m)
\end{align}
Functions $F_\pm(n)$ describe the effect of superposing a stationary qubit frequency noise on the effect of a periodic modulation of the qubit frequency. This effect is discussed in Sec.~\ref{sec:general_frequency_noise}. For the noise from TLSs and for Gaussian noise these functions have been essentially calculated in Ref.~\cite{Wudarski2023a}.

By construction, $F_+(n-m) \equiv F_+(|n-m|)$, whereas $F_-(n-m)=F^*_-(m-n)$. Also, we have $F_+(n)\to \langle \exp(i\theta^{(r)})\rangle^2$ and $F_-(n) \to |\langle\exp(i\theta^{(r)})\rangle|^2$ for $n\to \infty$. Here we use that, for a stationary noise, $\langle \exp(i\theta_k^{(r)})\rangle$ is independent of $k$, so that the subscript $k$ can be omitted,
\[\langle \exp(i\theta_k^{(r)})\rangle \to \langle \exp(i\theta^{(r)})\rangle. \]
For Gaussian noise $\langle \exp(i\theta^{(r)})\rangle$ is real (see below), and therefore the values of $F_+(n)$ and $F_-(n)$ coincide for $n\to \infty$.

The term $\mS_\mathrm{wht}$ describes the  noise that comes from the inherent randomness of the outcomes of quantum measurements. This noise provides a white-noise  background in the spectral measurements.

We show below that, as a consequence of the periodic modulation of the qubit frequency, the spectrum $\mS(m)$ displays resonant peaks. In particular, there emerge peaks for $m/N\approx \ell \omega_p\ts/2\pi$ with $\ell\geq 1$. To find these peaks we have to write the cosines in Eq.~(\ref{eq:power_synchron}) in terms of exponentials and, using the explicit form (\ref{eq:frequency_sinusoidal}) of $\theta_k^{(p)}$,  expand $\exp[\pm i \theta_{n_1}^{(p)}]$, $\exp[\pm i \theta_{n_2}^{(p)}]$ in series in $\exp(i n_{1} \omega_p\ts)$, $\exp(i n_{2} \omega_p\ts)$  \cite{Abramowitz1972}.  Equation~(\ref{eq:power_synchron}) then becomes a sum of terms  $\propto \exp[i\omega_p\ts (\ell_1n_1 + \ell_2 n_2) + 2\pi i m(n_1-n_2)/N]$. The spectral peaks are determined by the  terms with $\ell_2 = -\ell_1$ in this sum. To describe the peaks we approximate the expectation value of the product of the cosines in Eq.~(\ref{eq:power_synchron}) as 
\begin{align}
\label{eq:cosine_product}
&\braket{\cos(\theta_{n_1}+\phi_R) \cos(\theta_{n_2}+\phi_R)} \to C_\mathrm{res}(n_1,n_2),\nonumber\\
&C_\mathrm{res}(n_1,n_2)= \frac{1}{2}\sum_{\ell} J_\ell ^2(A_p) \mathrm{Re} \,\Bigl\{\exp[-i\ell(n_1-n_2)\omega_p\ts] 
\nonumber\\
&
\times\left[F_-(n_1 -n_2) + (-1)^\ell F_+(n_1-n_2)e^{ 2i\phi_R}\right]\Bigr\}
\end{align}
where $J_k(z)$ is the Bessel function of the first kind. We emphasize that the  expression for $\mS_\mathrm{syn}$ based on the approximation (\ref{eq:cosine_product})  applies only near the peaks of the spectrum. 

We will  calculate the spectrum $\mS_\mathrm{wht}$ assuming that the classical noise that leads to the random-phase term $\theta_k^{(r)}$ is stationary. We will also assume that $\omega_p\ts$ is not too close to a multiple of $2\pi$, so that $|1-\exp(ik\omega_p\ts)|\gg 1/N$ for all integer $k$. Then, in the limit of large $N$,  
\begin{align}
    \label{eq:white_Bessel}
    \mS_\mathrm{wht} =\frac{1}{8}\left[1 - J_0(2A_p) \mathrm{Re}\,\braket{\exp(2i\theta_n^{(r)} + 2i\phi_R)}\right].
\end{align}

The noise intensity $\mS_\mathrm{wht}\equiv \mS_\mathrm{wht}(m)$ is non-negative and is independent of $m$. Thus the noise $\mS_\mathrm{wht}(m)$ is white. We note that, in contrast to  the squared Bessel functions in the expressions for $\mS_\mathrm{syn}$, the expression (\ref{eq:white_Bessel}) for the white-noise intensity contains a term linear in the Bessel functions.


\subsection{Non-synchronized measurements}
\label{subsec:non_synchronized}

In deriving Eq.~(\ref{eq:power_synchron}) we assumed that the phase $\phi_p$ of the periodic modulation is known  and the measurements have been synchronized with the modulation. This could only be done if the modulation period were known.
Alternatively, to obtain the power spectrum, one repeatedly performs $N$ measurements to find $X(m)$, each time starting at the time, which is not synchronized with the (generally, unknown) periodic modulation.  Therefore the value of the modulation phase $\phi_p\, \mathrm{mod}\, 2\pi$, which is counted off from the instant when the measurements start, becomes effectively random. The averaging of measurement outcomes then incorporates  averaging both over the random part of the qubit phase $\theta_n^{(r)}$ and the modulation phase $\phi_p$. We denote the averaging over $\phi_p$  by $\langle\cdot\rangle_{\phi_p}$.  The expression for the power spectrum becomes  
\begin{align}
\label{eq:qubit_spectrum_periodic_general}
&\mS(m) = \langle |X(m)|^2\rangle_{\phi_p} = \mS_{\phi_p}(m) +
\mS_\mathrm{wht},\nonumber \\
&\mS_{\phi_p}(m)= 
 N^{-1}\sum_{n_1,n_2=0}^{N-1}e^{2\pi im(n_1 - n_2)/N}
\langle p(\theta_{n_1}) p(\theta_{n_2})\rangle_{\phi_p} .
\end{align}
Physically, the averaging over $\phi_p$ restores the symmetry with respect to the ``origin'' of time. Indeed,  the correlator $\langle p(\theta_{n_1}) p(\theta_{n_2})\rangle_{\phi_p}$ depends on $\theta_{n_1} - \theta_{n_2}$. Still  the symmetry with respect to time translation by the period $2\pi/\omega_p$ is in place. The form of the power spectrum reflects this symmetry by displaying sharp spectral peaks at the overtones of $\omega_p$. These peaks are described in Sec.~\ref{sec:resonant_peaks}.

Using Eq.~(\ref{eq:standard_probability}), we can write for $t_R\ll T_2$ 
\begin{align*}
&\langle p(\theta_{n_1}) p(\theta_{n_2})\rangle_{\phi_p} = [\langle p(\theta_{n_1})+ p(\theta_{n_2})\rangle_{\phi_p}]/2
\nonumber\\
& + \frac{1}{4}[\langle \cos(\theta_{n_1}+\phi_R) \cos(\theta_{n_2} + \phi_R)\rangle_{\phi_p} -1]
\end{align*}
As in the case of measurements synchronized with the modulation, only the averaged product of the cosines $\braket{\cos(\theta_{n_1}+\phi_R) \cos(\theta_{n_2} + \phi_R)}_{\phi_p}$ contributes to the spectrum $\mS_{\phi_p}(m)$ for $m\neq 0$. One immediately sees that this product is given by the function $C_\mathrm{res}(n_1,n_2)$ defined in Eq.~(\ref{eq:cosine_product}). This is not an approximation, in contrast to the case of synchronized measurements,
\begin{align}
    \label{eq:full_S_phi}
\mS_{\phi_p}(m) = \frac{1}{4N}\sum_{n_1,n_2=0}^{N-1}
C_\mathrm{res}(n_1,n_2) e^{2\pi i m(n_1 - n_2)/N}. 
\end{align}

The major contribution to the peaks of the power spectrum comes from the values $|n_1-n_2|\sim N$ which, for a long measurement sequence (large $N$), significantly exceed the correlation time of the frequency fluctuations $t_\mathrm{corr}$ divided by $\ts$. Therefore in calculating $\mS_{\phi_p}(m)$ near its peaks, to the leading order in $t_\mathrm{corr}/N\ts$ one may replace the functions $F_\pm(n_1-n_2)$ by their values for $|n_1-n_2|\to \infty$.


\section{Resonant peaks in the spectrum}
\label{sec:resonant_peaks}

The power spectrum we study has symmetry $\mS(m) = \mS(N-m)$. The low-frequency modulation and the low-frequency qubit noise  lead to the structure of $\mS(m)$ with typical width $\Delta m$ small compared to $N$, for  long data arrays of interest to us. Therefore we will consider $\mS(m)$ for $m<N/2$. Here the primary feature of the spectrum $\mS(m)$ is the occurrence of the periodic-modulation induced peaks at $m/N\approx \ell \omega_p \ts/2\pi$ with integer $\ell$. They come from the terms in Eq.~(\ref{eq:cosine_product}) with $\ell \geq 1$, and we assume $\ell \omega_p \ts/2\pi < 1/2$. From Eqs.~(\ref{eq:cosine_product}) and (\ref{eq:full_S_phi}), near an $\ell$th spectral peak we have
\[S_{\phi_p}(m) \approx \mS(m|\ell), \quad |(2\pi m/N) - \ell\omega_p\ts|\ll \omega_p\ts, 1,
\]
where
\begin{align}
\label{eq:spectral_peak}
&\mS(m|\ell)= 
 Q_\ell(m)\left\{|\braket{\exp[i\theta^{(r)}]}|^2\right. \nonumber\\
&\qquad \left.  +(-1)^\ell \mathrm{Re}\left[\braket{\exp[i\theta^{(r)}]}^2
\,\exp(2i\phi_R)\right]\right\}, \nonumber\\
&Q_\ell(m)= \frac{1}{8N} J_\ell^2(A_p)\frac{\sin^2(\ell N\omega_p \ts/2)}
{\sin^2[(2\pi m/N) - \ell\omega_p \ts)/2]}
\end{align}
The whole spectrum $\mS(m)$ is a superposition of the peaks (\ref{eq:spectral_peak}) complemented by the contribution from the frequency noise $\omq^{(r)}(t)$, see Sec.~\ref{sec:general_frequency_noise} below, and the quantum measurement noise $S_\mathrm{wht}$. The peaks can be associated with the overtones of the modulation frequency $\omega_p$, Their heights are $\propto N$, whereas their widths are $\propto 1/N$. 

In the limit of large $N$, where we can think of $2\pi m/N$ as a continuous variable, the $m$-dependent factor in Eq.~(\ref{eq:spectral_peak}) takes the form
\begin{align}
\label{eq:delta_peak}
&Q_\ell (m)\to  \frac{\pi}{4} J_\ell^2(A_p)\delta\left(\frac{2\pi m}{N} - \ell \omega_p\ts\right)
\end{align}

Generally, the spectrum $\mS(m)$ is even with respect to $m$, and in particular $\mS_{\phi_p}(m)$ has $\delta$-like peaks for positive and negative $m$, $\mS_{\phi_p}(m) = \mS_{\phi_p}(-m)$, as seen from Eq.~(\ref{eq:cosine_product}.) There are also replicas of the peaks, which are shifted by $N$ along the $m$-axis. For example, for $m>0$, there are peaks at $2\pi m/N\approx |\ell|\omega_p\ts$ and at $2\pi m/N\approx 2\pi -|\ell|\omega_p\ts$ with integer $|\ell|\geq 1$. In what follows we assume that the observation is long, $N\gg 1, |\ell|\omega_p\ts/2\pi$ for typical $\ell$, so that these peaks are well-separated. We consider the peaks with $0<m\ll N$. In terms of the comparison with the experiment, it is important that the factors $J_\ell(A_p)$ in Eq.~(\ref{eq:spectral_peak}) fall off fast with increasing $|\ell|$ for large $|\ell|$, so that even for a comparatively strong modulation of the qubit frequency the number of well-resolved peaks in the spectrum $\mS(m)$ is not large.

\begin{figure*}
    \centering
    \includegraphics[width = 0.97\textwidth]{ 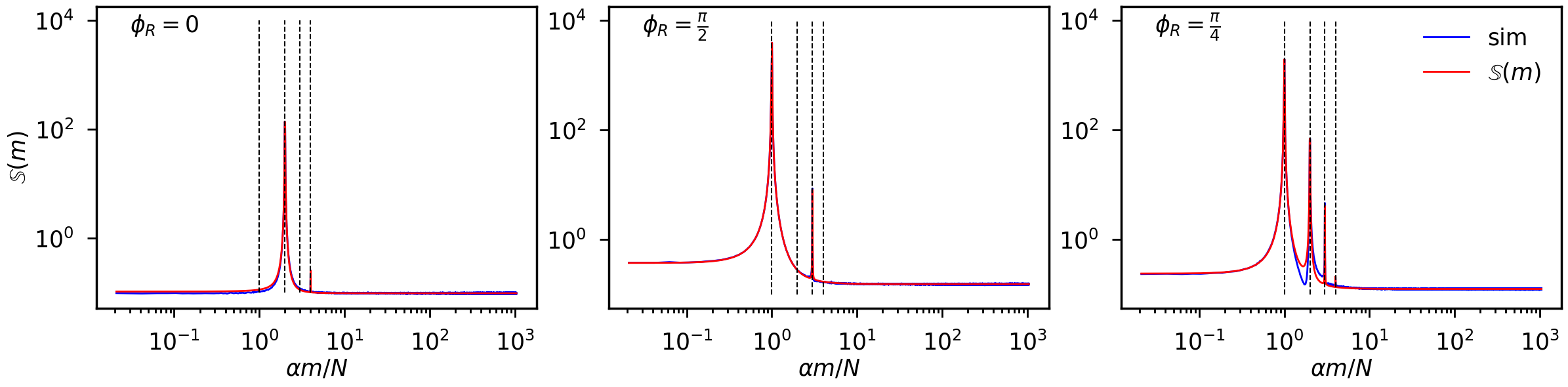}
    \caption{Comparison between the analytical expressions (\ref{eq:qubit_spectrum_periodic_general}) and (\ref{eq:spectral_peak}) for the power spectrum $\mS(m)$ (red lines) and  simulations (blue lines) for different values of the control phase $\phi_R$. The data are shown on the log-log scale. They refer to $\omega_p t_R = 10^{-3}$,  $\ts = 3 t_R$, and the modulation amplitude $a_p = t_R^{-1}$. The scaling parameter is $\alpha = 2\pi/\omega_p\ts$.  The dashed lines indicate the positions of the  four harmonics with the lowest values of $m/N>0$ in Eq.~(\ref{eq:spectral_peak}), i.e., $m/N = \ell \omega_p\ts/2\pi$ with $\ell = 1,2,3,4$. }
    \label{fig:periodic_only}
\end{figure*}

A characteristic feature of $\mS(m)$ that follow from Eq.~(\ref{eq:spectral_peak}) is that, for certain $\phi_R$, one would observe only even  or only odd overtones of $\omega_p$, i.e., the peaks at only even or only odd $\ell$. If one writes 
\begin{align}
    \label{eq:Theta_for_even_ell}
\braket{\exp[i\theta^{(r)}]}= \exp(-R+i\Theta^{(r)}),
\end{align}
then for $\phi_R=-\Theta^{(r)}$  only even overtones of $\omega_p$ will be observed, whereas for $\phi_R=-\Theta^{(r)} +\pi/2$ only odd overtones will be observed. For a zero-mean Gaussian noise or for the noise from symmetric TLSs $\Theta^{(r)}=0$, and then even or odd overtones are observed for $\phi_R=0$ or $\phi_R=\pi/2$, respectively. In a more general case, for example, for noise from asymmetric TLSs (see Sec.~\ref{subsubsec:TLSs}), the values of $\phi_R$ where the parity-dependent suppression occurs are shifted away from $0$ or $\pi/2$.

The onset of sharp peaks in the spectrum $\mS(m)$ in the presence of periodic modulation of the qubit frequency is seen in Fig.~\ref{fig:periodic_only}. The figure presents a comparison of the analytical results, Eqs.~\eqref{eq:qubit_spectrum_periodic_general} - \eqref{eq:delta_peak}, and numerical simulations. The simulation routine is described in Appendix~\ref{sec:appendix_simulations}. In particular, the simulations were done for $N=10^5$ and repeated  $10^4$ times for statistical averaging. The results refer to the case where random modulation of the qubit can be disregarded. 
We note that, in the absence of qubit frequency noise, 
\[\mS_{\phi_p}(m) = \sum_{|\ell|>0} \mS(m|\ell).\]
This expression was used in comparing the theory with the simulations.

Figure~\ref{fig:periodic_only} demonstrates that the number of well-resolved peaks of $\mS(m)$ depends on the value of the control phase $\phi_R$. For $\phi_R=0$ and $\phi_R = \pi/2$ visible are only the peaks at $m/N \approx \ell \omega_p\ts/2\pi$ with even and odd $\ell$, respectively, whereas for $\phi_R=\pi/4$ one can see peaks for all $\ell$.

\begin{figure}[h]
    \centering
    \includegraphics[width=0.47\textwidth]{ 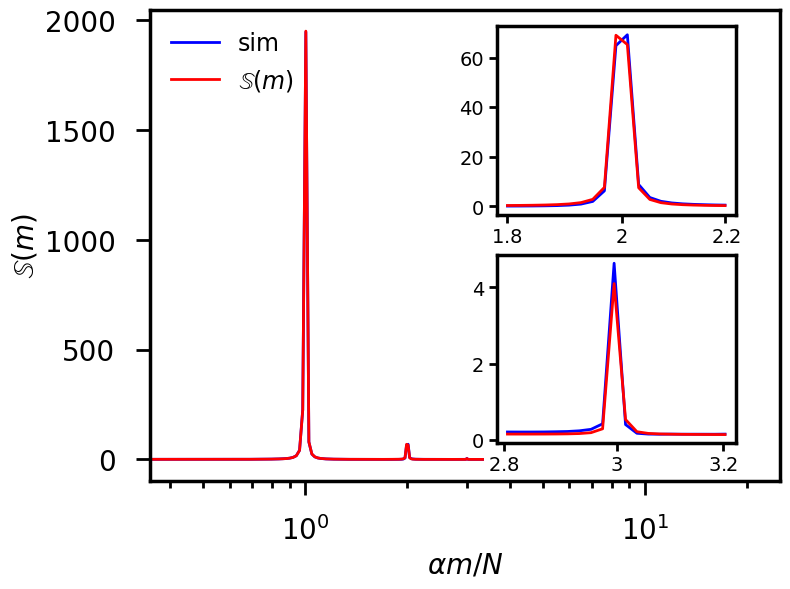}
    \caption{The power spectrum $\mS(m)$ on the linear scale for $\phi_R=\pi/4$. Other parameters are the same as in Fig.~\ref{fig:periodic_only}. The upper and lower insets show the peaks for $m/N \approx \ell \omega_p\ts/2\pi$ with $\ell = 2$ and $\ell=3$, respectively.   }
    \label{fig:periodic_only_lin}
\end{figure}

Figure~\ref{fig:periodic_only_lin} shows the spectrum $\mS(m)$ on the semi-log scale. On this scale the shape of the peaks is better visible. In interpreting the data one should keep in mind that the values of $m$ are discrete. The lines in the figure connect discrete dots. The maximum of the $\ell$th peak is $\propto N^{-1}\max_m [(2\pi m/N) - \ell\omega_p\ts]^{-2}$. 

The shapes of the spectral peaks are  different if only one integer $m$ is close to the resonant value $N\ell \omega_p\ts/2\pi$, as in the main plot and in the lower inset, or if two consecutive values of $m$ are close to $N\ell \omega_p\ts/2\pi$, as in the upper inset. In this latter case one can think of the peak as a doublet, with the partial peaks located at $m_> =  \lceil N\ell \omega_p \ts /2\pi\rceil$ and $m_<=  \lfloor N\ell \omega_p \ts /2\pi\rfloor  $. 

\subsection{Characterizing the frequency modulation from the power spectrum}
\label{subsec:interpolation}

One of the goals of studying the power spectrum $\mS(m)$ is to find  the parameters of the periodic modulation of the qubit frequency. The spectral peaks provide means to do this. To achieve high accuracy one has to use the values of $\mS(m)$ for several values of $m$, not just the peak value. An important information is provided by the tails of the peaks. It follows from Eq.~(\ref{eq:spectral_peak}) that these tails are described by a power law. For $2\pi m/N = \ell \omega_p\ts + \ep$ with $|\ep|\ll 1$, 
\begin{align}
    \label{eq:tail}
Q_\ell(m) \propto (4N)^{-1} J_\ell^2(A_p)/\ep^2.
\end{align}
This characteristic scaling should be identifiable in the data.

\begin{figure}[h]
    \centering
    \includegraphics[width=0.47\textwidth]{ 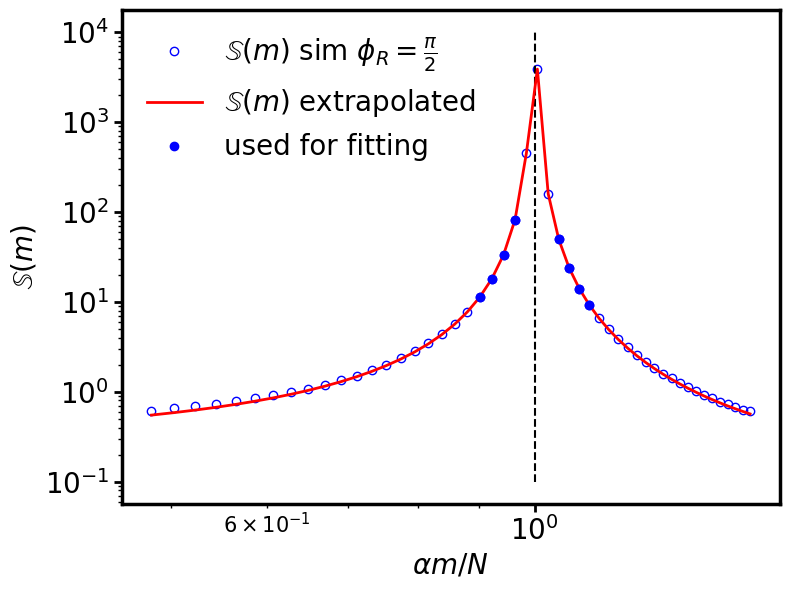}
    \caption{Finding the modulation parameters from the spectral peak of $\mS(m)$. The open dots show the results of simulations in the vicinity of the peak with $\ell=1$; these results were used in Fig.~\ref{fig:periodic_only}. The dashed line shows the value of $\omega_p\ts$ used in the simulations. The solid dots were used to fit $\mS(m)$ based on Eq.~(\ref{eq:spectral_peak}). The solid line shows the spectrum $\mS(m)$ calculated using the parameters found from the fit.}
    \label{fig:fitting}
\end{figure}

In Fig.~\ref{fig:fitting} we show how to use simulation points to determine the modulation parameters $\omega_p$ and  $A_p$. We use the data for the peak in Fig.~\ref{fig:periodic_only} for $\ell = 1$ and $\phi_R=\pi/2$. We fit four points on each  side of the peak using Eqs.~(\ref{eq:spectral_peak}) and (\ref{eq:tail}) adding a constant to $\mS(m)$ as a fitting parameter to describe the smooth background. We exclude the three points closest to  the peak, including the peak itself, since $\mS(m)$ is order of magnitude larger at these points as at the other points and varies extremely strongly from point to point. This has led to robust results with a standard fitting routine, with $\omega_p$ found with  almost perfect precision and $A_p$ found with an error $0.1\%$.

For weak and slow periodic frequency modulation, which is of significant interest for the experiment,
\[\omega_p t_R\ll 1, \qquad A_p\approx a_pt_R\ll 1.\]
In this case, $J_\ell(A_p)\approx (A_p/2)^\ell/\ell!$ to the leading  order in $A_p$. Then the leading-order contribution to the spectrum $\mS(m)$ comes from the term with $\ell = 1$ in Eq.~(\ref{eq:spectral_peak}) and only one peak is pronounced for $0<m< N-\omega_p\ts/2\pi$, with $Q_1(m)$ given by Eq.~(\ref{eq:spectral_peak}) in which $J_1^2(A_p)\approx (a_p t_R)^2/4$. However, for $\phi_R=-\Theta^{(r)}$, dominating in the spectrum will be a weaker peak at $2\pi m/N \approx 2\omega_p\ts$, with amplitude $\propto (a_pt_R)^4$.
%

\section{Effects  of the  qubit frequency noise}
\label{sec:general_frequency_noise}

Stationary fluctuations of the qubit frequency  lead to two major effects on the spectrum $\mS(m)$. They change the height of the spectral peaks (\ref{eq:spectral_peak}) and also lead to a generally smooth background in the spectrum. We will consider these effects separately


\subsection{Change of the height of the spectral peaks}
\label{subsec:Debye_Waller}

The change of  the height  (area) of the peaks (\ref{eq:spectral_peak})  due to stationary frequency fluctuations can be described by an exponential factor $\exp(-2R)$. The physics here is similar to the physics that underlies the Debye-Waller factor in the theory of spectral peaks of x-ray and neutron scattering in solids \cite{Girvin2019}. We will discuss this effect for two major sources of the fluctuations, the noise from TLSs and a Gaussian noise.

\subsubsection{Noise from dispersive coupling to TLSs}
\label{subsubsec:TLSs}

The value of the factor $\braket{\exp(i\theta^{(r)})}$ in Eq.~(\ref{eq:spectral_peak}) for the noise from TLSs was essentially obtained in \cite{Wudarski2023a}. In a simple model the TLSs are independent from each other and are described by the Pauli operators $\tau_z\sn$, where $n$ enumerates the TLSs. The TLSs are coupled to the qubit dispersively, the qubit frequency depends on the TLSs states, 
\begin{align}
\label{eq:zero_mean_TLS_noise}
\delta\omega_\mathrm{q}^{(r)}(t) = \sum_nV\sn{} (\tau_z\sn -\langle \tau_z\sn\rangle),
\end{align} 
where $V\sn$ is the parameter of the coupling to the $n$th TLS. 

The relevant TLSs dynamics are determined by the rates $W_{ij}\sn$ of switching $\ket{i}\sn\to \ket{j}\sn$ between their states, with $i,j$ taking values $0$ and $1$. The effect on the qubit is determined by the parameters  
\begin{align}
    \label{eq:TLS_switching_rates}
W\sn = W_{01}\sn + W_{10}\sn, \qquad \Delta W\sn = W_{10}\sn - W_{01}\sn.
\end{align}
In particular, $\Delta W\sn$ is the parameter of the TLS asymmetry, it shows the difference between the switching rates between different states and also determines the difference in the stationary state populations.

The switching occurs at random, leading to qubit frequency fluctuations. As a result of such fluctuations 
\begin{align}
\label{eq:phase_TLS}
\braket{\exp(i\theta^{(r)})} =  \prod_n e^{-iV\sn t_R\Delta W\sn/W\sn}\,\Xi\sn(t_R)
\end{align}
where
\begin{align}
\label{eq:Xi_n}
&\Xi\sn(t_R)=\Bigl[\left(\frac{W\sn{}}{2\gamma\sn} +iV\sn{} \frac{\Delta W\sn{}}
{\gamma\sn W\sn{}}\right)\sinh \gamma\sn t_R
\nonumber\\
&  +\cosh\gamma\sn t_R\Bigr]\exp(-W\sn{}t_R/2).
\end{align}
The expression for $\Xi\sn$ coincides with the previously obtained expression for the factor that describes decay of the qubit operators $\langle \sigma_\pm(t)\rangle$ due to the coupling to TLSs \cite{Paladino2002}. The form of $\Xi\sn$ is determined by the parameter $\gamma\sn$,
\begin{align}
\label{eq:mu_parameter}
&\gamma\sn=\frac{1}{2}\left[W\sn{}{}^2 +4iV\sn{} (\Delta W\sn{}+iV\sn{} )\right]^{1/2}, 
\end{align}
which depends on the interrelation between $V\sn{} $ and the TLS relaxation rate $W\sn$ given by Eq.~(\ref{eq:TLS_switching_rates}).

It follows from Eqs.~(\ref{eq:phase_TLS}) - (\ref{eq:mu_parameter}) that $\braket{\exp[i\theta^{(r)}]}$ is real if all TLSs are symmetric, $\Delta W\sn = 0, \forall n$. In this case in Eq.~(\ref{eq:Theta_for_even_ell}) the phase $\Theta^{(r)}=0$, as mentioned before. Then the spectral peak with $\ell=1$, which is of primary interest for weak periodic modulation, is pronounced for $\phi_R=\pi/2$. 

An important asymptotic of Eq.~(\ref{eq:phase_TLS}) is that of short Ramsey time, i.e., of small $|\theta^{(r)}|$. One can see that, in this case,
\begin{align}
\label{eq:small_theta_TLS}
&\braket{\exp[i\theta^{(r)}]} \approx 1 -\frac{1}{2}\sum_n w\sn{}^2V\sn{}^2 t_R^2 \nonumber\\
&+\frac{1}{3}\sum_n w\sn{}^2V\sn{}^2\left(\frac{W\sn}{2} + i\frac{V\sn \Delta W\sn}{W\sn}\right)t_R^3
\end{align}
where $w\sn = 2(W_{01}\sn W_{10}\sn)^{1/2}/W\sn$. Thus the imaginary part of   $\braket{\exp[i\theta^{(r)}]}$ emerges only in the 3rd order in $t_R$. Somewhat unexpectedly, it is of the first order in the TLS asymmetry $\Delta W\sn$.


\subsubsection{Gaussian noise of the qubit frequency}
\label{subsubsec:Gaussian}

It is easy to see that, if the frequency fluctuations $\delta\omega_q^{(r)}$ and thus the phases $\theta_k^{(r)}$, which are defined in terms of $\delta\omega_q^{(r)}$  by Eq.~(\ref{eq:phase_total}),  are Gaussian and zero-mean, then   
\begin{align}
\label{eq:Gaussian}
\braket{\exp[i\theta^{(r)}]} = \exp(-R)), \quad R=\braket{\theta^{(r)\,2}}/2.
\end{align}
Here we have taken into account that $\braket{\exp[i\theta^{(r)}]} \equiv \braket{\exp[i\theta_k^{(r)}]}$ is independent of $k$ for stationary noise, as mentioned earlier. Clearly, $\braket{\exp[i\theta^{(r)}]}$ is real.


\subsection{Noise-induced background of the power spectrum}
\label{subsec:background}

In addition to reducing the intensity of the spectral peaks of $\mS(m)$, the stationary noise that modulates the qubit frequency leads to the onset of a broad smooth contribution to the spectrum. A smooth spectrum exists in the absence of periodic modulation as well, but it is modified by the modulation. Formally, it is accounted for by the functions $F_\pm(n)$ in the general expressions for the spectrum (\ref{eq:power_synchron}). These functions have a particularly simple form for Gaussian noise, $F_\pm \to F_\pm^{(G)}$. A straightforward extension of the analysis of \cite{Wudarski2023a} shows that
\begin{align}
\label{eq:F_pm_Gauss}
&F_\pm^{(G)}(n_1 - n_2) = \exp(-2R \mp f_{n_1n_2}), \quad f_{n_1n_2} = \braket{\theta_{n_1}^{(r)}\theta_{n_2}^{(r)}} 
\end{align}
for $|n_1 - n_2|\geq 1$, where
\begin{align}
    \label{eq:f_nn}
&f_{n_1n_2} \approx \frac{1}{2\pi} t_R^2  \int d\omega S_q(\omega)\exp[-i\omega\ts (n_1 - n_2)],\nonumber\\
&S_q(\omega) = \int dt \exp(i\omega t)\braket{\omq^{(r)}(t)\omq^{(r)}(0)}.
\end{align}
Here $S_q(\omega)$ is the power spectrum of the frequency noise. Equation~(\ref{eq:f_nn}) is written for large $|n_2 - n_1|\ts \gg t_R$, where the main contribution to $f_{n_1n_2}$ comes from low-frequency fluctuations of $\omq^{(r)}$. In particular, it has been assumed that $\omega t_R\ll 1$ for typical values of $\omega$ that contribute to the integral over $\omega$ in the expression for $f_{n_1n_2}$. We also assumed that noise is classical and zero-mean, $\braket{\omq^{(r)}}=0$, and we used that, for stationary  noise, the correlator $\braket{\omq^{(r)}(t_1)\omq^{(r)}(t_2)}$ depends only on $t_1 - t_2$.

If the correlators $f_{n_1n_2}$ are small, 
\[F_\pm^{(G)}(n_1 - n_2) \approx e^{-2R}\left(1 \mp f_{n_1n_2}\right).\]
This expression can be extended to non-Gaussian noise. Quite generally, for large $|n_1-n_2|$ the correlation between $\theta_{n_1}^{(r)}$ and $\theta_{n_2}^{(r)}$ should be weak. Then  the noise correlators $f_{n_1n_2}$ defined by Eq.~(\ref{eq:f_nn}) will be small. From Eq.~(\ref{eq:phase_average}), for large $|n_1-n_2|$ the expansion of $F_\pm(n_1 - n_2)$ up to the term linear in $f_{n_1n_2}$ can be written as
\begin{align}
\label{eq:F_pm_approx}
F_\pm(n_1 - n_2) = \braket{e^{i\theta^{(r)}}}\braket{e^{\pm i\theta^{(r)}}}  \mp \zeta_\pm f_{n_1n_2},
\end{align}
where the factor $\zeta_\pm$ is determined by the noise statistics. This equation is an expansion in the noise correlation.  If the  noise is weak overall, i.e., for all $|n_1 - n_2|$, $\zeta_\pm = 1$; generally, $\zeta_- \geq |\zeta_+|$.  

It follows from Eqs.~(\ref{eq:cosine_product}), (\ref{eq:full_S_phi}), and (\ref{eq:F_pm_approx}) that, in the presence of periodic modulation of the qubit frequency  and for $N\gg 1$, the background of $\mS(m)$ induced by low-frequency stationary noise has the form 
\begin{align}
\label{eq:background_ell}
&\Delta \mS(m) = \frac{t_R^2}{8\ts}\sum_\ell J_\ell^2(A_p)\left[\zeta_- - (-1)^\ell \mathrm{Re}\,\zeta_+e^{2i\phi_R}\right]\nonumber\\
&\times S_q\left(\frac{2\pi m}{N\ts} - \ell\omega_p\right) 
\end{align} 
The sum over $\ell$ here runs  from $\ell\to -\infty$ to $\ell \to \infty$. Again, one can set $\zeta_\pm=1$ for weak noise correlations. 

Equation (\ref{eq:background_ell})  shows that the spectrum $\Delta \mS(m)$ is given by a sum of the spectra of the stationary qubit frequency noise $S_q(\omega)$ at frequencies shifted by the multiples of the frequency $\omega_p$ of the periodic modulation. The term with $\ell=0$ gives the spectrum in the absence of periodic  modulation multiplied by a factor $J_0^2(A_p)$, which reduces its height. 

The total spectrum in the presence of periodic and random modulation of the qubit frequency is 
\begin{align}
    \label{eq:total_spectrum}
    \mS(m) \approx \sum_{\ell>0} \mS(m|\ell) + \Delta\mS(m) + S_\mathrm{wht}.
\end{align}
If the width of the frequency noise spectrum $S_q(\omega)$ is small compared to $\omega_p$, Eq.~(\ref{eq:background_ell}) describes a background on which there is located an $\ell$th $\delta$-like peak  $\mS(m|\ell)$ described by Eq.~(\ref{eq:spectral_peak})]. Conversely, if the spectrum $S_q(\omega)$ is broad, the backgrounds of different peaks of $\mS(m|\ell)$ overlap. It is important that the peaks themselves are not broadened, they remain $\delta$-like.

\begin{figure}[t]
    \centering
    \includegraphics[width=0.47\textwidth]{ 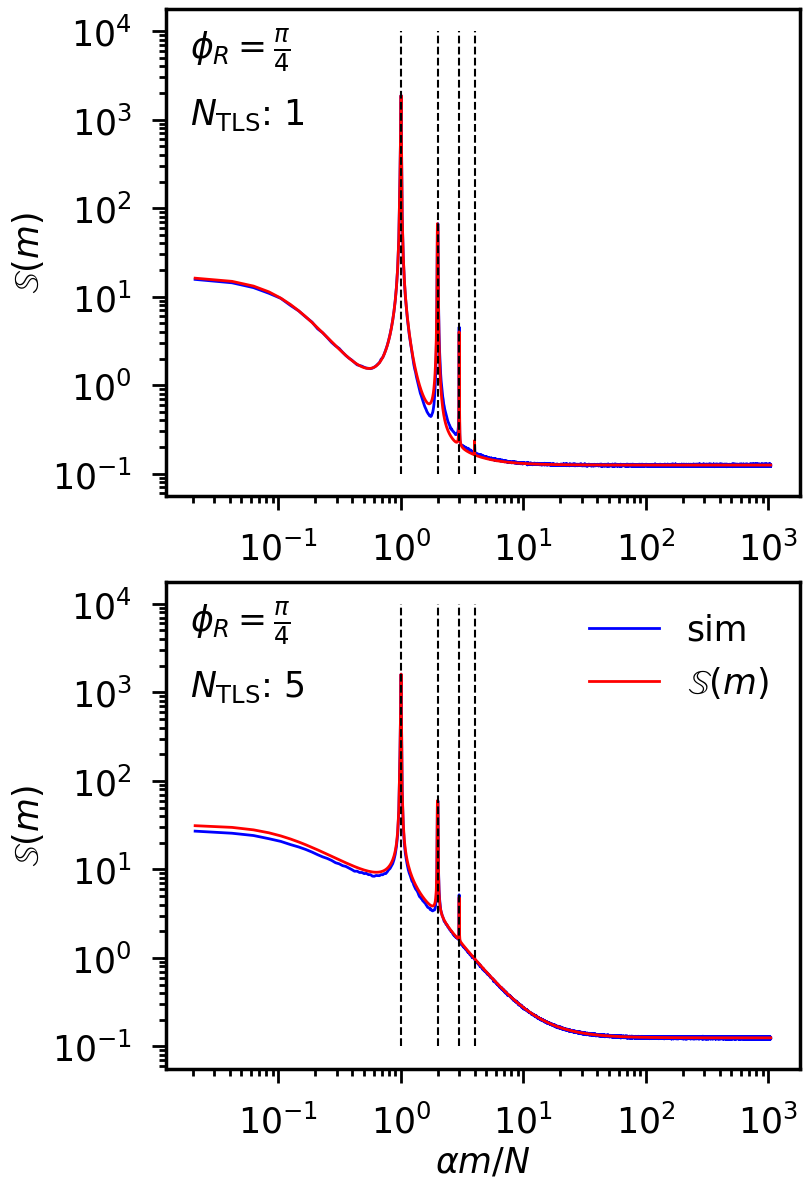}
    \caption{Power spectrum $\mS(m)$ in the presence of periodic and random modulation of the qubit frequency. The parameters of the periodic modulation are the same as in Fig.~\ref{fig:periodic_only}. The upper panel refers to the random modulation from the coupling to a single TLS, $N_\mathrm{TLS}=1$, whereas the lower panel refers to $N_\mathrm{TLS}=5$. The dispersive coupling strength is the same for all TLSs, $V_n=0.2 t_R^{-1}$. The TLSs are symmetric. The switching rate of the single TLS is $W^{(1)} = 1.2\times 10^{-4}t_R^{-1}$, whereas for the five TLSs $W^{(n)} = \exp\left(-3 n/4\right)t_R^{-1}$ with $n=8, 9,...,12$. The red and blue lines show the analytical result (\ref{eq:total_spectrum}) and the results of the simulations, respectively. The dashed lines indicate the values of $m/N = \ell \omega_p\ts/2\pi$ with $\ell=1,2,3,4$.}
    \label{fig:TLS}
\end{figure}

The combined effect of periodic and random  modulation of the qubit frequency is illustrated in Fig.~\ref{fig:TLS}. The parameters of the periodic modulation are the same as in Fig.~\ref{fig:periodic_only}, whereas the frequency noise was assumed to come from a different number $N_\mathrm{TLS}$ of symmetric TLSs. For $N_\mathrm{TLS}=1$ the noise is a telegraph noise, with $S_q(\omega) \propto V_1^2 W^{(1)}/(W^{(1)\,2} + \omega^2)$. For the chosen switching rate $W^{(1)}$, the width of this spectrum is smaller than the distance between the periodic-modulation induced peaks of $\mS(m)$, so that these peaks are superposed on the smooth background $\Delta \mS$. 

For the values of $W^{(n)}$ chosen in the lower panel of Fig.~\ref{fig:TLS}, the noise $S_q(\omega)$ from 5 TLSs mimics $1/f^2$ noise over more than five decades. Here the relative heights of the peaks of $\mS(m)$ change, as the width of $S_q(\omega)$ is no longer small compared to $\omega_p$. As seen from the figure, the spectrum $\mS(m)$ is very different from what would come out if there were superposed the spectrum from periodic modulation $\sum_\ell \mS(m|\ell)$ and the spectrum $\mS(m)$ from the TLS-induced noise in the absence of periodic modulation. Nontrivial mixing of the contributions of different mechanisms of the qubit frequency modulation is a feature of such modulation. 

\section{Effect of breaking the time- translation symmetry}
\label{sec:peak_broadening}

The periodicity of the qubit frequency modulation $\omega_p$ along with the periodicity of the measurements $\ts$ impose discrete time-translation symmetry on the outcomes of the Ramsey measurements. Breaking this periodicity restores continuous time-translation symmetry. In turn, this leads to the broadening of the spectral peaks (\ref{eq:spectral_peak}).

\subsection{Effect of fluctuations of the modulation frequency}
\label{subsec:omega_p_fluctuations}

We start with the analysis of frequency fluctuations, as they are unavoidably present in any signal, making it quasiperiodic. We will assume that the fluctuations are slow compared to $\omega_p$. With the account taken of such fluctuations, the quasiperiodic modulation of the qubit frequency is described by the expression  
\begin{align}
\label{eq:quasiper_modulation}
\omq^{(qp)}(t) = a_p\cos\left[\omega_p t+ \int_0^t dt'\xi(t')\right].
\end{align}
Here $\xi(t)$ is the noise of the modulation frequency. We assume this noise to be small,  $\sigma[\xi]\ll \omega_p$, with $\sigma[\xi]$ being the standard deviation of $\xi(t)$, and smooth on the time scale $\omega_p^{-1}$. Physically, the major effect of the noise $\xi(t)$ is the random phase that is accumulated over a long sequence of measurements. It is this phase that  leads to a broadening of the spectral peak.

To calculate the effect, we start with the phase acquired by the qubit during a $k$th Ramsey measurement. To the leading order in $\xi(t)$
\begin{align}
\label{eq:theta_qp}
&\theta_k^{(qp)} = \int_{k\ts}^{k\ts + t_R}\omq^{(qp)}(t)dt \approx A_{p}\cos(k\omega_p\ts + \Phi_k+\tilde\phi_p),
\nonumber\\
&A_p\approx \frac{2a_p}{\omega_p}\sin\frac{\omega_p t_R}{2}, \quad \Phi_k = \int_0^{k\ts} dt \,\xi(t),
\end{align}
where 
\[\tilde\phi_p \approx (\omega_p t_R/2) + \frac{1}{2}\int_{k\ts}^{k\ts+t_R}\xi(t) dt.\] 
It would be more accurate to replace $\omega_p$ with $\omega_p + \xi(k\ts)$ in the expression for $A_p$; however, the corresponding small corrections do not accumulate during repeated measurements. Moreover, in the interesting case $\omega_p t_R\ll 1$ the frequency $\omega_p$ drops out from the expression for $A_p$. We will also disregard the small random term $\sim \sigma[\xi] t_R \ll 1$ in $\tilde\phi_p$.

Because of the accumulation of frequency fluctuations, the factor $\exp[-i\ell (n_1 - n_2)\omega_p\ts]$ in Eq.~(\ref{eq:cosine_product}), which is responsible for the peaks of $\mS(m)$ at  $2\pi m/N=\ell\omega_p \ts$,  is multiplied by 
\begin{align}
\label{eq:Xi_factor}
&\varrho_{n_1,n_2}=\mE\bigl[\exp[-i\ell (\Phi_{n_1}-\Phi_{n_2})]\bigr]\nonumber\\
&=\mE\left[\exp\bigl[-i\ell \int_{n_1\ts}^{n_2\ts}dt\, \xi(t)\bigr]\right].
\end{align}

The factor $\varrho$ can be calculated for Gaussian fluctuations of the modulation frequency, with the result expressed in term of the noise correlator $\Xi(t)$,
\begin{align}
    \label{eq:xi_correlator}
\braket{\xi(t_1)\xi(t_2)} = \Xi(t_1 - t_2), \qquad \Xi(t)=\Xi(-t).
\end{align}
In the important case where the correlation time of the fluctuations is small compared to  the total measurement duration, we have 
\begin{align} 
\label{eq:expon_decay}
&\varrho_{n_1n_2} \approx \exp\left[-\Gamma_\ell |n_1 - n_2|\right],\nonumber\\
&\Gamma_\ell=\frac{1}{2}\ell^2 \ts  \int_{-\infty}^\infty dt \,\Xi(t).
\end{align}

The exponential decay of the factor $\varrho_{n_1n_2}$ with the increasing $|n_1-n_2|$ leads to broadening of the peaks in the spectra of the measurement outcomes. In particular, in the limit of large $N$, the peaks $Q_\ell$ in Eq.~(\ref{eq:delta_peak}), instead of being $\delta$-functions, become Lorentzians,
\begin{align}
\label{eq:Lorentzians}
&Q_\ell (m)\to  \frac{1}{4} J_\ell^2(A_p) \Gamma_\ell \bigg/\left[
\Gamma_\ell^2 + \left(\frac{2\pi m}{N} - \ell \omega_p\ts\right)^2\right]. \end{align}
The factor $\Gamma_\ell$ gives the half-widths of the spectral peaks, but the overall area of the peaks, $\sum_m Q_\ell(m)$, does not change. $\Gamma_\ell$  is linear in the variance of the fluctuations of the modulation frequency. It quickly increases with the number $\ell$ of the overtone of $\omega_p$. It also determines the heights of the Lorentzian peaks, which are now $\propto 1/\Gamma_\ell$ and are independent of $N$.

\begin{figure}[h]
    \centering
    \includegraphics[width = 0.47\textwidth]{ 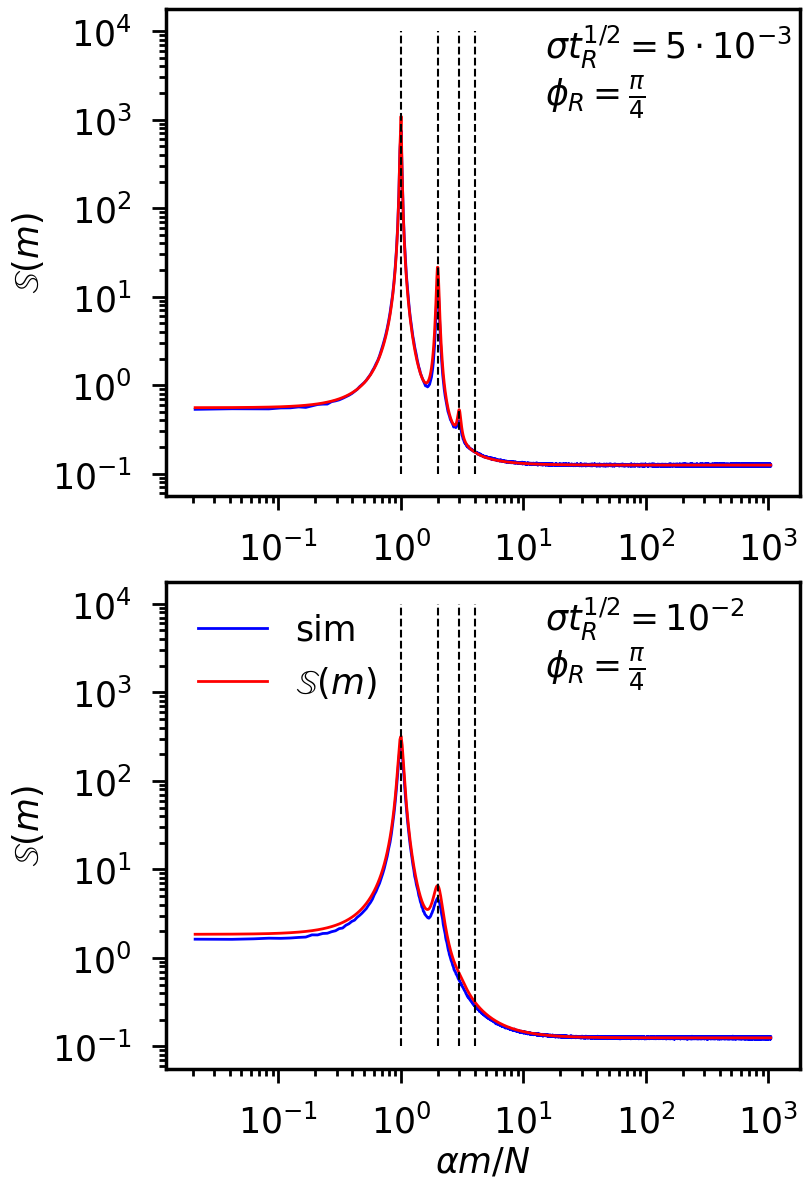}
    \caption{The power spectrum $\mS(m)$ in the presence of fluctuations of the modulation frequency. The fluctuations are described by white Gaussian noise with intensity $\sigma^2$ defined in Eq.~(\ref{eq:sigma_modulation_noise}). The upper and lower panels refer to $\sigma t_R^{1/2} = 5 \,\omega_p t_R$ and $10\, \omega_pt_R$, respectively. The other parameters are the same as in Fig.~\ref{fig:periodic_only}. The red and blue lines show the theory, Eqs.~\eqref{eq:Lorentzians} and \eqref{eq:spectral_peak}, and the results of simulations. The dashed lines show the positions of the peaks in the absence of fluctuations, $m/N = \ell\omega_p\ts/2\pi$ with $\ell = 1,2,3,4$. }
    \label{fig:white}
\end{figure}

The spectrum $\mS(m)$ in the presence of fluctuations of the modulation frequency is illustrated in Fig.~\ref{fig:white}. The plots refer to a delta-correlated Gaussian noise $\xi(t)$ with two values of the intensity $\sigma^2$, which are given by the relation
\begin{align}
    \label{eq:sigma_modulation_noise}
\Xi(t) \equiv \braket{\xi(t)\xi(0)}= \sigma^2\delta(t)
\end{align}
(see Appendix~\ref{sec:appendix_simulations}  for the details of the simulations). For such noise the halfwidths of the spectral peaks of $\mS(m)$ are $\Gamma_\ell = \frac{1}{2} \ell^2 \ts \sigma^2$. In the upper panel of Fig.~\ref{fig:white}, where $\sigma^2=25 \omega_p^2 t_R$, the halfwidths $\Gamma_\ell$ are smaller than  the interpeak distance $\omega_p\ts/2\pi$ for $\ell = 1,2,3$ and $\omega_p t_R=10^{-3}$, so that the peaks with $\ell = 1,2,3$  are  resolved in the spectrum. In the lower panel of Fig.~\ref{fig:white}, where $\sigma^2 = 10^2 \omega_p^2 t_R$, the peaks with $\ell>2$ cannot be resolved. It is seen from the figure that the heights of the peaks fast decreasing with the increasing $\ell$, in agreement with the theory. Appendix~\ref{sec:appendix_simulations} shows that a similar evolution of the spectrum occurs for exponentially correlated Gaussian noise as the noise correlation time increases.


\subsection{Fluctuations of the cycle duration}
\label{subsec:t_c_fluctuations}

An important role in the broadening of the spectral peaks can be played by fluctuations of the cycle duration. Such fluctuations occur where  $\ts$ varies from cycle to cycle. The duration of the $k$th cycle is then $\ts^{(k)} = \ts + \delta \ts^{(k)}$, where $\ts\equiv \braket{\ts^{(k)}}$ is the mean duration of the cycle,  $\braket{\delta\ts^{(k)}} = 0$. In the presence of such fluctuations the qubit phase accumulated over the $k$th cycle is
\begin{align}
    \label{eq:random_t_cyc}
\theta_k =A_p\cos[(k\ts + \tau_k)\omega_p +\tilde\phi_p], \quad
\tau_k = \sum_{n=0}^k \delta\ts^{(n)}
\end{align}
with $A_p$ and $\tilde \phi_p$ given by Eq.~(\ref{eq:frequency_sinusoidal}). We do not consider here the contribution from random modulation of the qubit frequency discussed in the previous subsection, its effect is not different from that in the absence of fluctuations of $\ts$.

As in the analysis of the case where the frequency $\omega_p$ was fluctuating, we see that 
the factor $\exp[-i\ell (n_1 - n_2)\omega_p\ts]$ in Eq.~(\ref{eq:cosine_product}), which is responsible for the peaks at frequencies $2\pi m/N=\ell\omega_p \ts$,  is multiplied by 
\begin{align}
\label{eq:delta_ts_factor}
&\tilde\varrho_{n_1 n_2}=\mE\bigl[\exp[-i\ell \omega_p(\tau_{n_1}-\tau_{n_2})]\bigr]
\end{align}

The factor $\tilde\varrho$ is easy to calculate for Gaussian fluctuations of $\ts^{(k)}$. If the correlator of $\delta\ts^{(k)}$ is
\[\mathcal{T}(n_1 - n_2) = \braket{\delta\ts^{(n_1)}\delta\ts^{(n_2)}},\quad \mathcal{T}(n_1 - n_2)= \mathcal{T}(n_2 - n_1),\]
we have 
\begin{align*}
    \tilde \rho_{n_1n_2} =& \exp\Bigl[ - \frac{1}{2}\ell^2\omega_p^2\sum_{k=-(|n_1-n_2|-1)}^{|n_1-n_2|-1} (|n_1-n_2|-|k|)\nonumber\\
 &\times   \mathcal{T}(k)\Bigr].
    \end{align*}
In the important case where $(n_1 - n_2)t_{\rm cyc}$ is much larger than the correlation time of $\delta t_{\rm cyc}$, this expression simplifies to 

\begin{align}
    \label{eq:tilde_varrho}
&\tilde\varrho_{n_1n_2} \approx \exp\left[-\tilde\Gamma_\ell |n_1 - n_2|\right], \nonumber\\
&
\tilde\Gamma_\ell =\frac{1}{2}\ell^2\omega_p^2\sum_{k=-\infty}^\infty \mathcal{T}(k).
\end{align}
The above expression for $\tilde\varrho_{n_1n_2}$ is exact if $\delta\ts^{(k)}$ are uncorrelated for different $k$.

The change  of the spectral peaks $\mS(m|\ell)$ in the presence of fluctuations of the period of the cycle is similar to that in the presence of  fluctuations  of the modulation frequency $\omega_p$. The expression for $Q_\ell(m)$ is of the form of  Eq.~(\ref{eq:Lorentzians}) with $\Gamma_\ell$ replaced by $\tilde\Gamma_\ell$.

\section{Conclusions}
\label{sec:conclusions}

This paper provides a framework for revealing periodic modulation of the qubit frequency and characterizing it. 
Of primary interest and importance is the modulation at frequencies much lower than the coherence time of the qubits. As shown in the paper, such modulation leads to the onset of sharp peaks in the spectrum of the outcomes of periodically repeated Ramsey measurements. The positions of the peaks are  determined by the product of the modulation frequency and the period at which the measurements are repeated. 

Because of the qubit nonlinearity, the spectrum can display peaks at several overtones of the modulation frequency. Unexpectedly, different overtones are displayed depending on  the phase $\phi_R$ that is accumulated during a Ramsey measurement because of the  difference between the frequencies of the qubit  and the reference signal.

It is important that the positions and the sharpness of the peaks are not affected by the presence of low-frequency qubit noise if the noise is not too strong. This allows finding the modulation frequency and amplitude with high accuracy by using the shapes and the positions of the peaks. The low-frequency qubit noise leads to a smooth background in the spectrum. However, this background is modified by the periodic modulation of the qubit frequency. The spectrum is not a superposition of the spectrum in the absence of periodic modulation and the modulation-induced peaks.

Fluctuations of the modulation frequency or the periodicity of the measurements break the time-translation symmetry of the system and the measurement procedure. In turn, this leads to broadening of the spectral peaks. For the both mechanisms the broadening increases quadratically with the overtone number. The widths of the peaks are proportional to the intensity of the corresponding fluctuations. When the fluctuations are not too strong, the widths of the peaks are small compared to the inter-peak distance, and then the mean modulation frequency is well-defined and can be found from the measurements.

The analytical results of the paper are in excellent quantitative agreement with the results of extensive simulations of the qubit dynamics. These combined results suggest a benchmarking protocol applicable to various quantum computing platforms. Identifying the frequency of the modulation should help establishing the fluctuation source and thus improving qubit quality.

\begin{acknowledgments}
We are grateful to A. N. Korotkov, who participated in the work at the early stage. FW and MID acknowledge partial support from NASA Academic Mission Services, Contract No. NNA16BD14C, and from Google under NASA-Google SAA2-403512. The research of MID was supported in part by the Google Research Award.
    
\end{acknowledgments}



\appendix

\section{SIMULATIONS}
\label{sec:appendix_simulations}

We test the developed theory in simulations, which allows us to identify features that might be elusive from the theoretical description. In particular, we focus on three main types of results: i) a qubit with the frequency subjected to periodic modulation only; here the qubit dynamics is tested for different values of $\phi_R$, ii) a qubit which is additionally coupled to TLSs, and iii) the case where the modulation frequency is fluctuating. The results of the simulations are displayed in figures in the main text. In the following sections we provide description of the simulation setups.

\subsection{Simulation of periodic modulation}
Since the modulation-induced phase accumulation is deterministic, see Eq.~\eqref{eq:frequency_sinusoidal}, and depends on setup's parameters only, one can easily determine the effect  of the periodic modulation on the qubit dynamics and then on the measurement outcomes. We simulate the sequence of $N$ periodically repeated Ramsey measurements by setting up the phases $\theta^{(p)}_k$ accumulated in each measurement,  $k=1,2,\ldots, N$, using Eq.~\eqref{eq:frequency_sinusoidal}. In our simulation we use $N=10^5$ outcomes. For each $k$ we have certain probability $p(\theta_k)$ of observing outcome `1' according to Eq.~\eqref{eq:standard_probability} (we operate in fully coherent regime, i.e. $T_2\to\infty$). Next, for each $p(\theta_k)$ we sample a number $r_k$ from a uniform distribution $U(0,1)$. If $p(\theta_k) \ge r_k$, we assign outcome `1', otherwise `0'. The entire process of collecting $N$ measurement outcomes is repeated $K=10^4$ times to collect statistically relevant outcomes. The obtained results are then used to calculate discrete Fourier transform Eq.~\eqref{eq:fast_Fourier_general} that determines the power spectrum. 

Throughout all simulations we set the Ramsey accumulation time $t_R=1$ (arb. units) and the duration of the entire cycle $\ts=3t_R$. The modulation frequency is $\omega_pt_R=0.001$ and the modulation amplitude is $a_p=1$. We test different phases $\phi_R$ that are reported in all figures. Finally, we set the initial phase of the modulation to be $\phi_p=0$, as it provides the same results as selecting it at random from a uniform distribution $U(0,2\pi)$ if averaged over many $K$ repetitions, which has been verified in the simulations. 

In Fig.~\ref{fig:fitting} we show how to use simulation points to determine $\omega_p$ and  $A_p$. As described in the main text, in order to achieve that we use Eq.~\eqref{eq:spectral_peak} for $\ell=1$ (i.e. for the first peak) as the fitting function  with addition of a free parameter $c$ to capture the offset stemming from the measurement noise (the background).
For the fitting procedure we use a {\it trf} method available in \texttt{scipy.optimize.curve\_fit} \cite{virtanen2020scipy} (we use SciPy 1.11.1), for which we specify the boundary conditions for fitting parameters. The boundary conditions can be estimated directly from the spectrum with accuracy $1/N$ for $\omega_p$ and $A_p$, where the latter is determined by integrating the peak. This approach allows us to find $\omega_p$ with almost perfect accuracy for this setup and $A_p$ with $0.1\%$ accuracy.

\subsection{Simulations of a qubit subjected to periodic frequency modulation and the noise from  TLSs}
Simulations of the qubit frequency  subjected to  periodic modulation and to an additional noise  can be performed independently. We consider noise from coupling the qubit to a set of TLSs. One can simulate the frequency noise first as described in \cite{Wudarski2023a} (code available in \cite{Wudarski2022}), and then simulate periodic modulation as presented in the previous subsection. The resulting  frequencies are then added together, the phase in the $k$-th Ramsey measurement is calculated using  Eq.~\eqref{eq:phase_total}, and the result is substituted into  the probability distribution of the measurement outcomes. This allows finding the  measurement outcomes based on uniformly sampling  numbers from $[0,1]$ interval.

\subsection{Noise in the frequency $\omega_p$ of periodic modulation}

Finally, we test the influence of the noise $\xi(t)$ added to  the periodic modulation frequency $\omega_p$. We consider two types of noise: (i) Gaussian white noise and (ii) exponentially correlated Gaussian noise. Simulations of the both cases are based on Eq.~\eqref{eq:quasiper_modulation}. In the case (i), $\xi(t)$ is drawn from a normal distribution centered at $\mu=0$ and with standard deviation $\sigma$, i.e. from the distribution $\mathcal{N}(\mu,\sigma^2)$. In the case (ii), $\xi(t)$ is given by the Ornstein-Uhlenbeck (OU) process and is described by the stochastic differential equation
\begin{equation}\label{eq:OU_process}
    d\xi(t) = - \frac{\xi(t)dt}{\tau_{\mathrm{corr}}} + \sqrt{\frac{2D}{\tau_{\mathrm{corr}}}}dW(t),
\end{equation}
where $W(t)$ denotes Wiener process. Since noise in $\omega_p$ is time-dependent, simulations require full time-evolution, which is done through
discretization into increments of $dt = 0.1t_R$ and $dt=0.01t_R$ for Gaussian white and exponentially correlated noise, respectively. In the case of Gaussian white noise, we simulate the integral by sampling at each time step $dt$ a number according to the distribution $\sigma \sqrt{dt} \mathcal{N}(0,1)$. While for the OU process we use the following representation

\begin{figure}[h]
    \centering
    \includegraphics[width = 0.47\textwidth]{ 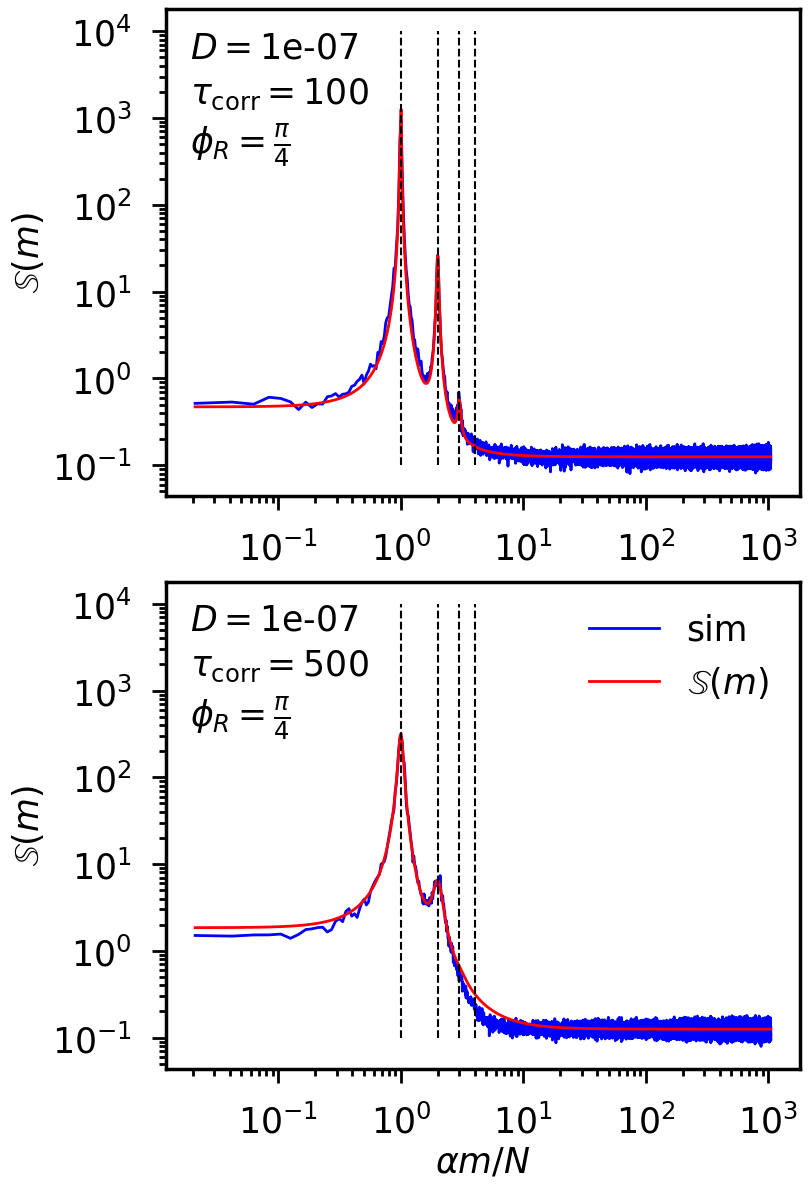}
    \caption{Comparison between theory based on Eqs.~\eqref{eq:Lorentzians} and \eqref{eq:spectral_peak} and simulation of a system subjected to exponentially correlated noise with noise intensity $D$ and correlation time $\tau_{\mathrm{corr}}$ present in $\omega_p$. The other parameters are the same as in Fig.~\ref{fig:periodic_only}}
    \label{fig:exp_corr}
\end{figure}
\begin{equation}
    \xi(t + dt) = \xi(t) -\frac{\xi(t)dt}{\tau_{\mathrm{corr}}} + \sqrt{\frac{2 D}{\tau_{\mathrm{corr}}}} \mathcal{N}(0,\sqrt{dt}).
\end{equation}
At the next step,  $\xi(t)$ is numerically integrated with the same $dt$ discretization. With the both methods, we  determine the increment  $\delta\omega_q^{(qp)}(t)$ over  the entire duration of the Ramsey cycles. These increments  are then integrated to obtain $\theta_k^{(qp)}$ for each $k$-th measurement in the presence of modulation frequency noise. Subsequently, we determine the outcomes of the measurements as in the noiseless scenario. Since, simulation of the OU process is computationally demanding, we restricted the simulations to $K=100$ repetitions of the entire measurement process, still keeping $N=10^5$ outcomes in each of them.

The results for Gaussian white noise are depicted in Fig.~\ref{fig:white} and are described in the main text.
For the exponentially correlated case (see Fig.~\ref{fig:exp_corr}) we see similar behavior as in the case of Gaussian white noise. However, the shape of the spectrum is in this case controlled by the product  $\tau_{\mathrm{corr}}D$, since the noise correlation function is $\Xi(t) = D\exp(-|t|/\tau_\mathrm{corr})$, while for white noise $\Xi(t)=\sigma^2\delta(t)$.

\section{MEASURING MODULATION FREQUENCY USING TUNABLE FOURIER TRANSFORM}
\label{sec:frequency_tuning}

An alternative way to determine the frequency of the frequency-modulation drive is based on tuning the frequency of the discrete Fourier transform of the measurement outcomes. Using the  array of the outcomes $\{x_n\}$ one calculates function 
\begin{align}
    \label{eq:omega_transform}
    Y(\nu; M) = \frac{1}{N}\sum_{n=0}^{M-1} \exp(i\nu n)x_n
\end{align}
Of interest are large $M\gg 1$ while still $M<N$. The goal is to use $Y(\nu;M)$ to measure the modulation frequency $\omega_p$. The idea is that the dependence of $Y(\nu;M)$ on $M$ is very sensitive to the detuning of $\nu$ from $\omega_p\ts$, and therefore by adjusting $\nu$ one can find $\omega_p$ with high precision. 

We will first consider the case where the qubit frequency is modulated only by the periodic drive, so that the total qubit phase at the $n$th measurement is $\theta_n
=\theta_n^{(p)} = A_p\cos(n\omega_p\ts+ \tilde\phi_p)$. For low-frequency modulation we have $\omega_p\ts\ll 1$. Then the phase $\theta_n$ changes weakly for many consecutive $n$, and for $M\gg 1$ and $\nu\sim \omega_p \ts$ we have
\begin{align}
    \label{eq:Y_to_mean_Y}
Y(\nu; M)\approx \mE[Y(\nu;M)].
\end{align}
To calculate $Y(\nu;M)$ using this expression, we have to replace $x_n$ in Eq.~(\ref{eq:omega_transform}) with $\mE[x_n] = p(\theta_n) = [1+ \cos(\theta_n+\phi_R)]/2$. We can further express  $\cos(\theta_n+\phi_R)$ in terms of a series in powers of $\exp(in\omega_p\ts)$. After summing over $n$  we find for $\nu\approx \omega_p\ts$ 
\begin{align}
    \label{eq:Y_no_noise}
    |Y(\nu;M)|\approx \left\vert
    \frac{\sin\phi_R}{2N} J_1(A_p)\frac{\sin[(\nu-\omega_p\ts)M/2]}{{\sin[(\nu-\omega_p\ts)/2]}}
    \right\vert
\end{align}
Function $|Y(\nu;M)|$ is $\propto M$ for exact resonance, $\nu=\omega_p\ts$. This is a characteristic behavior. Deviations from the proportionality  $|Y(\nu;M)|\propto M$ occur for $|\nu-\omega_p\ts|\lesssim 1/M$. This shows that, for large $M$, the frequency $\omega_p$ can be found very precisely by tuning $\nu$ and studying the dependence of $Y(\nu;M)$ on $M$.

We now consider the effect of fluctuations of the driving frequency $\omega_p$, that is, the case where $\omega_p$ has a random term $\xi(t)$, $\omega_p\to \omega_p+ \xi(t)$. As before, we assume the noise $\xi(t)$ to be small, $\sigma[\xi]\ll \omega_p$, where $\sigma[\xi]$ is the standard deviation of $\xi(t)$. The phase of the drive accumulated over $n$ measurement cycles, along with $n\omega_p\ts$ acquires an extra random term $\Phi_n=\int_0^{n\ts} dt\,\xi(t)$, cf. Eq.~(\ref{eq:theta_qp}). This term remains nearly constant over many $n$ for a weak noise, and therefore for $\omega_p\ts\ll 1$ the phase of the qubit also remains nearly constant. As a consequence, function $Y(\nu;M)$ is self-averaging for large $M$ and can be calculated using Eq.~(\ref{eq:Y_to_mean_Y}). For a stationary Gaussian noise $\xi(t)$ with correlator $\Xi(t)$, see Eq.~(\ref{eq:xi_correlator}), we have
\begin{align}
&\braket{\exp\left(i\ell \Phi_n\right)} 
=\exp\left[-\ell^2\int_0^{n\ts} dt (n\ts-t)\Xi(t)\right]\nonumber
\end{align}

We will assume that the correlation time of the noise $\xi(t)$ is small compared to $M\ts$. Then for typical $n\sim M$ in the above expression $n\ts\gg t$ and we have $\braket{\exp\left(i\ell \Phi_n\right)}\approx \exp(-n\Gamma_\ell)$, where $\Gamma_\ell$ is given by Eq.~(\ref{eq:expon_decay}). As a result for $|\nu-\omega_p\ts|\ll 1$ we have 
\begin{widetext}

\begin{figure}
   \centering
  \includegraphics[width = 0.98\textwidth]{ 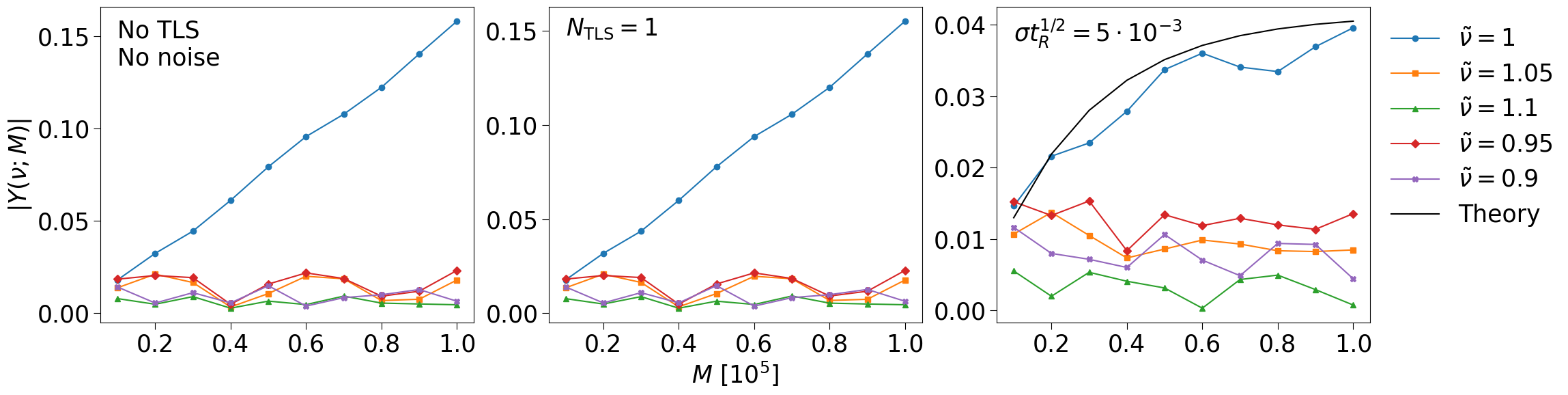}
 \caption{Finding modulation frequency $\omega_p$ with the tunable Fourier transform. Shown is the amplitude of the Fourier transform of the measurement outcomes $|Y(\nu;M)|$ as a function of the transform frequency $\nu$ and the number of measurements $M$; the value of $\nu$ is scaled by $\omega_p\ts$, i.e., $\tilde\nu = \nu/\omega_p\ts$; $\phi_R=\pi/4$. The parameters of the periodic modulation are the same as in Fig.~\ref{fig:periodic_only}. The left panel shows $|Y(\nu;M)|$ where the qubit frequency is changed only by the periodic modulation. The central panel show the effect of incorporating noise from a single TLS with the same parameters as in Fig.~\ref{fig:TLS}. The right panel shows the effect of white noise in the frequency of periodic modulation, with the noise parameters being the same as in Fig.~\ref{fig:white}.}
    \label{fig:freq_tuning}
\end{figure}
\end{widetext}
\begin{align}
    \label{eq:Y_with_freq_noise}
&|Y(\nu;M)|\approx \left|
\frac{\sin\phi_R}{2N} J_1(A_p) \right| \nonumber\\
&\times \{ 1+e^{-2\Gamma_1 M} -2e^{-\Gamma_1 M}\cos[M(\nu-\omega_p\ts)]\}^{1/2}\nonumber\\
&\times [\Gamma_1^2 + (\nu - \omega_p\ts)^2]^{-1/2}
\end{align}
For $\Gamma_1M\ll 1$ this expression coincides with Eq.~(\ref{eq:Y_no_noise}). Therefore $|Y(\nu;M)|$ displays  characteristic linear increase with $M$ on exact resonance, $\omega_p\ts = \nu$. However, for larger $M$ this increase saturates. For exact resonance we have 
\[|Y(\nu;M)|\propto (1-e^{-\Gamma_1 M})/\Gamma_1 \quad (\nu=\omega_p\ts)
\]
This expression clearly shows the saturation for $M\sim 1/\Gamma_1$. An advantageous feature of Eq.~(\ref{eq:Y_with_freq_noise}) is that not only it allows finding the modulation frequency $\omega_p$, but also the intensity of the fluctuations of this frequency.

A similar analysis can be repeated for the overtones, $\nu = \ell\omega_p\ts$ with $\ell>1$. We note that $\Gamma_\ell$ increases as $\ell^2$ with the increasing $\ell$, which should enable testing the fluctuation mechanism. The analysis of the effect of fluctuations of the cycle duration $\ts$ is also similar, with $\Gamma_\ell$ replaced by $\tilde\Gamma_\ell$.

In Fig.~\ref{fig:freq_tuning} we show the results of simulations of the spectral  amplitude $|Y(\nu;M)$ for several values of the trial frequency $\nu$. In the absence of extra noise, $|Y(\nu;M)|$ displays the expected linear increase with $M$ provided $\nu=\omega_p\ts$. If the trial frequency is detuned from exact resonance, the response changes dramatically, even a $\pm 5\%$ change in $\nu$ makes $|Y(\nu;M)|$ weakly dependent on $M$. 

The central panel of Fig.~\ref{fig:freq_tuning} shows that an extra weak noise in the {\it qubit} frequency does not qualitatively change $|Y(\nu;M)|$. This agrees with the previous observation that a weak qubit noise does not prevent finding the modulation frequency with high precision from the sharp peaks of the standard power spectrum $\mS(m)$.       

The right panel of Fig.~\ref{fig:freq_tuning} shows that noise in the modulation frequency significantly changes $|Y(\nu;M)$ even where the noise is weak. Besides the characteristic saturation with the increasing $M$ expected from Eq.~(\ref{eq:Y_with_freq_noise}), the data shows large spread of the simulated values of $|Y(\nu;M)|$ where $\nu$ is detuned from the average modulation frequency. On the one hand, this suggests a way of identifying weak nose of the modulation frequency using the proposed method but, on the other hand, it shows that the method is less efficient for finding the mean frequency itself.


\bibliography{refs}
\end{document}